%% file: paper.tex
\documentclass[10pt,conference]{IEEEtran}

\usepackage{graphicx}
\usepackage{tabularx}
\usepackage{colortbl}
\usepackage{xcolor}
\usepackage{amsfonts}
\usepackage{pifont}% http://ctan.org/pkg/pifont
\newcommand{\cmark}{\ding{51}}%
\newcommand{\xmark}{\ding{55}}%
\usepackage{tikz}

\usepackage{dingbat}
\usepackage{booktabs}
\usepackage{adjustbox}
\usepackage{listings}   %% code listings
\usepackage{xcolor}     %% code listings
\usepackage{subcaption} %% code listings
\usepackage{hyperref}
\usepackage{xurl}

\usepackage[colorinlistoftodos,prependcaption,textsize=tiny]{todonotes}

\definecolor{pred}{RGB}{204,51,51}

\definecolor{codeKey}{RGB}{050, 100, 200}
\definecolor{codeStr}{RGB}{220, 075, 150}
\definecolor{codeBgr}{RGB}{255, 255, 253}
\definecolor{codeCmt}{RGB}{050, 150, 050}
\definecolor{codeNmr}{RGB}{200, 200, 200}

\lstdefinestyle{listingStyle}{
    backgroundcolor=\color{codeBgr},
    commentstyle=\color{codeCmt},
    keywordstyle=\color{codeKey},
    numberstyle=\color{codeNmr},
    stringstyle=\color{codeStr},
    basicstyle=\ttfamily\footnotesize,
    breakatwhitespace=false,
    breaklines=true,
    captionpos=b,
    keepspaces=true,
    numbers=left,
    numbersep=5pt,
    showspaces=false,
    showstringspaces=true,
    showtabs=false,
    tabsize=4,
    frame=single,
    rulecolor=\color{black}
}

\newcommand{\listing}[1]{
    \begin{figure}[h]
    \centering
    \begin{subfigure}{0.9\linewidth}
        \lstinputlisting[language=Python]{cwe-#1p.py}
        \vspace{-3mm}
        \caption{Prompt}
    \end{subfigure}
    \begin{subfigure}{0.9\linewidth}
        \lstinputlisting[language=Python]{cwe-#1r.py}
        \vspace{-3mm}
        \caption{Copilot's Suggestion}
    \end{subfigure}
    \caption{Scenario CWE #1}
    \label{fig:#1}
    \vspace{-5mm}
    \end{figure}
}

\AtBeginDocument{%
  \providecommand\BibTeX{{%
    \normalfont B\kern-0.5em{\scshape i\kern-0.25em b}\kern-0.8em\TeX}}}

\begin{document}

\title{Assessing the Security of GitHub Copilot’s Generated Code - A Targeted Replication Study}

\author{\IEEEauthorblockN{Vahid Majdinasab\IEEEauthorrefmark{1},
Michael Joshua Bishop\IEEEauthorrefmark{2},
Shawn Rasheed\IEEEauthorrefmark{3}, Arghavan Moradidakhel\IEEEauthorrefmark{1},
Amjed Tahir\IEEEauthorrefmark{2}, Foutse Khomh\IEEEauthorrefmark{1}\\ \newline}
\IEEEauthorblockA{\IEEEauthorrefmark{1}Department of Computer and Software Engineering \\ Polytechnique Montreal, Canada\\
\IEEEauthorrefmark{2}School of Mathematical and Computational Sciences \\ Massey University, New Zealand\\
\IEEEauthorrefmark{3}Information \& Communication Technology Group \\ UCOL - Te P\=ukenga, New Zealand\\
Email: \IEEEauthorrefmark{1}\{arghavan.moradi-dakhe;vahid.majdinasa;foutse.khomh\}@polymtl.ca,
\IEEEauthorrefmark{2}a.tahir@massey.ac.nz,
\IEEEauthorrefmark{3}s.rasheed@ucol.ac.nz}}

% \author{
% \IEEEauthorblockA{\IEEEauthorrefmark{1}Department of Computer and Software Engineering \\ Polytechnique Montreal Canada\\}
% \and
% \IEEEauthorblockA{\IEEEauthorrefmark{2}School of Mathematical and Computational Sciences \\ Massey University, New Zealand\\}
% \and
% \IEEEauthorblockA{\IEEEauthorrefmark{3}Information \& Communication Technology Group \\ UCOL - Te P\=ukenga, New Zealand\\}
% }

% Email: \IEEEauthorrefmark{1}\{arghavan.moradi-dakhe;vahid.majdinasa;foutse.khomh\}@polymtl.ca,
% \IEEEauthorrefmark{2}a.tahir@massey.ac.nz,
% \IEEEauthorrefmark{3}s.rasheed@ucol.ac.nz}}

\maketitle

\begin{abstract}
AI-powered code generation models have been developing rapidly, allowing developers to expedite code generation and thus improve their productivity.
%AI-powered code generation models have been rapidly advanced, allowing developers to expedite code generation and thus improve their productivity.
These models are trained on large corpora of code (primarily sourced from public repositories), %(mostly hosted on public repositories),
which may contain bugs and vulnerabilities. Several concerns have been raised about the security of the code generated by these models. %The security of the generated code has been questioned.
Recent studies have investigated security issues in AI-powered code generation tools such as GitHub Copilot and Amazon CodeWhisperer, revealing %and reported
several security weaknesses in the code generated by these tools. As these tools evolve, it is expected that they will improve their security protocols to prevent the suggestion of insecure code to developers. %so they avoid suggesting insecure code to developers.
This paper replicates the study of Pearce et al., which investigated security weaknesses in Copilot and uncovered several weaknesses in the code suggested by Copilot across diverse scenarios and languages (Python, C and Verilog).
%In this paper, we replicate the first study that investigated security weaknesses in Copilot by Pearce et al., which uncovered several security weaknesses in code suggested by Copilot for prompts within various coding scenarios in multiple languages (Python, C, and Verilog).
Our replication examines Copilot’s security weaknesses using newer versions of Copilot and CodeQL (the security analysis framework). The replication focused on the presence of security vulnerabilities in Python code. Our results indicate that, even with the improvements in newer versions of Copilot, the percentage of vulnerable code suggestions has reduced from 36.54\% to 27.25\%. Nonetheless, it remains evident that the model still suggests insecure code.
\end{abstract}


\input{introduction}

\input{originalstudy}
\input{methodology}
\input{results}
\input{discussion}
\input{related-work}
\input{conclusion}

\input{acknowledgements}
\bibliographystyle{ieeetr}
\bibliography{paper.bib}

\end{document}

%% file: introduction.tex
\section{Introduction}

Code generation tools aim to increase productivity by generating code segments for developers - either in the form of auto-completion of existing code segments or converting prompts (written in natural language) into code. Code generation tools have been around for some time.
New code generation tools based on AI models, in particular, have gained popularity over the past few years with the availability of commercial products such as GitHub Copilot\footnote{\url{https://github.com/features/copilot}} and Amazon CodeWhisperer\footnote{\url{https://aws.amazon.com/codewhisperer}}. This includes the use of large language models (LLMs) that translate natural language into code. Such tools are touted as AI-pair programmers, trained on billions of existing lines of code that help write code (in multiple languages) faster and with less work by turning natural language prompts into coding suggestions. 

Copilot is based on models built using OpenAI’s Codex \cite{chen2021evaluating}, which interprets comments in natural language and executes them on the user’s behalf. The Copilot model is trained on publicly available code from projects hosted on GitHub.  
As of June 2022, Copilot has been used by more than a million developers and generated over three billion accepted lines of code \cite{ThomasGithub2023}.

Previous research has studied code generation tools, with more focus on the correctness of the results~\cite{lertbanjongngam2022empirical, wong2022exploring, dakhel2023github, pudari2023copilot}. There is also increased attention given to the security of the generated code~\cite{siddiq2022securityeval,he2023controlling}, including studies on new tools such as Copilot, CodeWhisperer, and ChatGPT \cite{yeticstiren2023evaluating,asare2023github}. 

Generating code based on the training on publicly available code may result in code that inherits not just the intended functionality or behavior but also bugs and security issues. Previous studies have shown that publicly available code, such as code snippets hosted on Stack Overflow, can be vulnerable \cite{verdi2020empirical}. This code leads to not only generating `functional' code but also security bugs and vulnerabilities. In the context of Copilot, the tool may produce insecure code as Codex, its model, is trained on code hosted on GitHub \cite{brown2020language}, which is known to contain buggy programs and untrusted data \cite{rokon2020sourcefinder}. 

In 2022, Pearce et al. \cite{Pearce22} studied security weaknesses of Copilot-generated code for several programming languages (i.e., Python and C)  and reported that 40\% of the generated programs contained vulnerabilities. The examples were generated using MITRE’s Common Weakness Enumerations (CWEs), from their ``2021 CWE Top 25 Most Dangerous Software Weaknesses". A recent study on the security weaknesses in Copilot-generated code found in publicly available GitHub projects (using multiple languages)
shows that over 35\% of Copilot-generated code snippets contain CWEs. It also reported the security weaknesses are diverse in nature and related to 42 different CWEs (from MITRE's list) \cite{fu2023security} (including CWEs that appear in MITRE's 2022 list). These findings confirm that such weaknesses can also make their way into real-world projects if generated code is not appropriately checked. Copilot security concerns go beyond vulnerable code suggestions; Copilot was found to reveal hard-coded secrets that were part of the training data in GitHub- a recent study \cite{huang2023not} found over 2,000 hard-coded credentials in Copilot-generated code, raising alarms of major privacy concerns of the potential leakage of hard-coded credentials. This is mainly because the GitHub training data also contains millions of hard-coded secrets \cite{YesGitHu93}. 

With the continued improvement in Copilot, it is expected that security measures will be put into place to filter out vulnerable code (that may introduce CWEs). We aim to test this by replicating the study of Pearce et al. using a variety of CWEs and prompts (as in the original study). The goal of this study is to conduct a targeted replication study of Copilot Python code using MITRE’s top 25 CWEs. \\
This replication study addresses two main questions: 
\textbf{does Copilot provide insecure code suggestions?} and \textbf{what is the prevalence of insecure generated code?}. We used Copilot to generate code suggestions using prompts based on 12 CWEs from MITRE's CWE Top 25 Most Dangerous Software Weaknesses.

Our results show that despite improvements in Copilot's newer versions in terms of filtering out vulnerable suggestions, it is still evident that many Copilot suggestions contain CWEs. This is the case across a diversity of weaknesses and scenarios. For the investigated Python suggestions, we found evidence of improvements, with a reduced number of vulnerable Copilot suggestions from 35.35\% to 25.06\%. We also noted that there are 100\% improvements with regard to some of the scenarios and CWEs (that is, the replication shows no weaknesses in the code suggestions). Interestingly, we note some cases where the new version of Copilot's suggestions contains more weaknesses than what has been shown in the original study. This shows that while Copilot has improved in filtering out vulnerable code suggestions, it does not completely eliminate them. Developers should be cautious of the code suggestions generated by Copilot. Developers should incorporate automatic and manual security analysis of the code before integrating Copilot suggestions into the code.

The paper is organized as follows: we explain the replication scope and our methodology is presented in Section \ref{sec:methodology}. We present our results in Section \ref{sec:results} followed by a discussion of these results in Section \ref{sec:discussion}. We present results work in Section \ref{sec:relatedwork}. Finally, Section \ref{sec:conclusion} presents the study conclusion and future research directions.

%% file: originalstudy.tex
\section{Original study}

The authors of the original study use Copilot with code prompts to answer these questions: 
Are Copilot’s suggestions commonly insecure? What is the
prevalence of insecure generated code? What factors of the
``context'' yield generated code that is more or less secure?
% \subsection{Study setup}
The original study examines Copilot's behavior across three dimensions: diversity of weakness, diversity of prompt, and diversity of domain. In this replication, we focus on just the diversity of the weakness dimension. The original study constructs three scenarios for each of ``top 25'' CWE's and uses CodeQL or manual inspection to determine security issues present in the generated code. For all axes and languages, 39.33\%  of the top and 40.73\% of the total options were vulnerable. For Python specifically, this number is 37.93\% of the top and 36.54\% of the total.

%% file: methodology.tex
\section{Replication scope and methodology}
\label{sec:methodology}

Table \ref{tab:setup} provides an overview of the experimental setups followed in the original study and in this replication. Details of our methodology for generating the CWE scenarios and our manual analysis are provided below.

\begin{table*}[]
\centering
\caption{Experimental setup in the original study and the replication}
\label{tab:setup}
 \begin{adjustbox}{max width=\textwidth}
\begin{tabular}{lll}
\toprule
Criteria & Original Study & Replication Study \\ \midrule
Scope & Python, C and Verilog & Python \\
Copilot version and date & \begin{tabular}[c]{@{}l@{}}Copilot technical preview version \\ (specific version number not  provided)\end{tabular}& 1.77.922 \\
CodeQL version &  2.5.7 &   2.12.4 \\
\# of CWEs & 12 & 12 \\
\# of scenarios & 54 & 29 \\ \bottomrule

\end{tabular}%
\end{adjustbox}
\end{table*}

% List CWEs investigated here
\begin{table*}[]
\caption{List of CWEs examined~\cite{CWE}}
\label{tab:cwes}
  \begin{adjustbox}{max width=\textwidth}

\begin{tabular}{@{}lll@{}}
\toprule
CWE \# & Name & Description \\ \midrule
20 & Improper Input Validation & \begin{tabular}[c]{@{}l@{}}The product receives input or data, but it does not validate or incorrectly validates that \\ the input has the properties that are required to process the data safely and correctly.\end{tabular} \\ \midrule
22 & Unauthorized Path Traversal & \begin{tabular}[c]{@{}l@{}}The product uses external input to construct a pathname that is intended to identify \\ a file or directory that is located underneath a restricted parent directory, but the \\ product does not properly neutralize special elements within the pathname that can \\ cause the pathname to resolve to a location that is outside of the restricted directory.\end{tabular} \\ \midrule
78 & OS Command Injection & \begin{tabular}[c]{@{}l@{}}The product constructs all or part of an OS command using externally-influenced input \\ from an upstream component, but it does not neutralize or incorrectly neutralizes \\ special elements that could modify the intended OS command when it is sent to a \\ downstream component.\end{tabular} \\ \midrule
79 & Cross-Site Scripting & \begin{tabular}[c]{@{}l@{}}The product does not neutralize or incorrectly neutralizes user-controllable input before \\ it is placed in output that is used as a web page that is served to other users.\end{tabular} \\ \midrule
89 & SQL Injection & \begin{tabular}[c]{@{}l@{}}The product constructs all or part of an SQL command using externally-influenced input \\ from an upstream component, but it does not neutralize or incorrectly neutralizes \\ special elements that could modify the intended SQL command when it is sent to a \\ downstream component.\end{tabular} \\
\midrule
200 & Exposure of Sensitive Information to an Unauthorized Actor & \begin{tabular}[c]{@{}l@{}}The product exposes sensitive information to an actor that is not explicitly authorized to\\have access to that information.\end{tabular} \\ \midrule
306 & Missing Authentication for Critical Function & \begin{tabular}[c]{@{}l@{}}The product does not perform any authentication for functionality that requires a provable\\user identity or consumes a significant amount of resources. \end{tabular} \\  \midrule
434 & Unrestricted Upload of File with
Dangerous Type & \begin{tabular}[c]{@{}l@{}}The product allows the attacker to upload or transfer files of dangerous types that can be \\automatically processed within the product's environment. \end{tabular} \\ \midrule
502 & Deserialization of Untrusted Data & \begin{tabular}[c]{@{}l@{}}The product deserializes untrusted data without sufficiently verifying that the resulting \\ data will be valid.\end{tabular} \\ \midrule
522 & Insufficiently Protected Credentials & \begin{tabular}[c]{@{}l@{}}The product transmits or stores authentication credentials, but it uses an insecure method\\that is susceptible to unauthorized interception and/or retrieval.\end{tabular} \\ \midrule
732 & Incorrect Permission on Critical Resource & \begin{tabular}[c]{@{}l@{}}The product specifies permissions for a security-critical resource in a way that allows \\ that resource to be read or modified by unintended actors.\end{tabular} \\ \midrule
798 & Use of Hard-Coded Credentials & \begin{tabular}[c]{@{}l@{}}The product contains hard-coded credentials, such as a password or cryptographic key, \\ which it uses for its own inbound authentication, outbound communication to external \\ components, or encryption of internal data.\end{tabular} \\ \bottomrule
\end{tabular}
\end{adjustbox}
\end{table*}

\subsection{Generating CWE scenarios using Copilot}

In this study, we focus our experimental analysis on the Python language and analyzed scenarios related to twelve CWEs (details are shown in Table~\ref{tab:cwes}).
For each CWE, we adapted the scenarios from the original study of Pearce et al. ~\cite{Pearce22}. The original study employed three distinct sources to develop these scenarios. The initial source includes CodeQL examples and documentation. The second source encompasses examples provided for each CWE in MITRE's dataset. The final source involves scenarios designed by the author.

Copilot is prompted to complete each scenario by placing the cursor at the position where code completion is needed. It is important to note that these scenarios do not contain vulnerabilities; our objective is to examine whether the code added by Copilot to complete the scenarios introduces any vulnerabilities.

Our objective was to gather 25 solutions proposed by Copilot for each scenario. Although Copilot's configuration in Visual Studio Code IDE \footnote{\url{https://code.visualstudio.com}} provides various settings, such as the option to define the maximum number of displayed solutions (\textit{ListCount}), in the version (1.77.9225) utilized for this study, this parameter does not function as intended. Regardless of our attempts to set the maximum solutions anywhere from 10 to 100, the tool consistently returns a random quantity of solutions in each iteration. To address this limitation, we collect up to 55 solutions generated by Copilot for each scenario across multiple iterations.

Given that this study aims to investigate potential vulnerabilities in Copilot's suggestions, our evaluation concentrates solely on identifying vulnerabilities within the provided suggestions rather than assessing their accuracy in addressing the prompts. However, it is important to note that some of the suggestions proposed by Copilot exhibit syntax errors. We excluded such suggestions from our evaluation process for two reasons. First, in order to scan them for security weaknesses using CodeQL, the code must be compilable. Moreover, in line with the original study \cite{Pearce22} and the work of Dakhel et al. \cite{dakhel2023github}, there are potential instances of duplication in Copilot's top-n suggestions. To address this issue, we adopted the same strategy as recommended in~\cite{dakhel2023github} to eliminate duplicates. This approach involves computing the similarity between the Abstract Syntax Trees (ASTs) of different suggestions, with an exclusion of the leaves corresponding to variable or function names (including all natural language content). In order to compare the ASTs of solutions, the codes must not contain any syntax errors and therefore we exclude those that cannot be compiled.

For removing the duplicate solutions, in cases where the similarity between two ASTs is identified as 1, we categorize them as duplicates and subsequently eliminate one of them. It is noteworthy that, in the original study~\cite{Pearce22}, duplicates were removed by comparing the sequences of tokens, which resulted in the identification of distinct duplicate code snippets in their collected solutions, differing only in variable or function names or content of comments in different lines of codes. Using AST similarity to remove the duplicates between the suggested solutions allows us to have a more accurate understanding of Copilot's faults. %This is why we only focus on the Python programming language, as comparing the ASTs of programs in this language is more dependable than C and Verilog.

% because, in order to assess their security vulnerabilities using CodeQL, the code must be compilable.
% \todo[inline]{@Vahid add a note about discarded solutions here }

% Furthermore, in line with the work of Dakhel et al. ~\cite{dakhel2023github}, there are potential instances of duplication in Copilot's top-n suggestions. To address this issue, we adopted the same strategy as recommended in~\cite{dakhel2023github} to eliminate duplicates. This approach involves computing the similarity between the Abstract Syntax Trees (ASTs) of different suggestions, with an exclusion of the leaves corresponding to variable or function names (including all natural language content). In cases where the similarity between two ASTs is identified as 1, we categorize them as duplicate suggestions and subsequently eliminate one of the duplicates. It's noteworthy that, in the replication study~\cite{Pearce22}, duplicates were removed by comparing the sequences of tokens, which resulted in the identification of distinct duplicate code snippets in their collected solutions, differing only in variable or function names or content of comments in different lines of codes. Using AST similarity to remove the duplicates between the suggested solutions allows us to have a more accurate understanding of Copilot's faults. Therefore, we only focus on the Python programming language as comparing the ASTs of programs in this language is more dependable than C and Verilog.

\subsection{Static Analysis with CodeQL}
\begin{figure*}[h]
\centering
\includegraphics[width=0.7\linewidth]{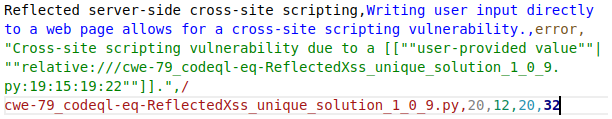}
\caption{Example CodeQL output for Copilot-generated Python Code}
\label{fig:codeql-out}
\end{figure*}

CodeQL analysis was conducted on a machine with an \texttt{Intel Xeon E5-2690 v3} processor with a memory of \texttt{128GB DDR4 RAM } running \texttt{Ubuntu 20.04.5 LTS}. The analysis took 16 minutes 37 seconds real-time. % or an equivalent of 53 minutes 48 seconds CPU time.
 Automated Python scripts were used to run the database build and analysis tests based on parameters specific to each sample set/CWE. All scenarios and scripts used during our analysis were developed for Python 3.8.10. CodeQL version 2.12.4 of the command line toolchain was used for static security analysis. An example of CodeQL's output for a vulnerability is shown in Fig. \ref{fig:codeql-out}.

% \todo[]{add python version for code generation as opposed to analysis}

\subsection{Manual Analysis}
% \todo[inline]{maybe also add a note on the reasons for this manual analysis of the selected samples?}

We conducted a manual analysis for the code suggestions that we could not analyze using CodeQL. We analyzed a total of 141 code samples related to four CWEs (i.e., CWE-434, CWE-306, CWE-200 and CWE-522).
This manual analysis was conducted by two coders (both are co-authors). Both coders have experience with security analysis and have discussed the vulnerabilities and the review protocol before conducting the analysis. Each coder was assigned a set of code suggestions to validate whether they contained the specific CWE under investigation.
The two coders first analyzed 12 code suggestions (3 suggestions for each of the four CWEs) to verify their classification and agree on the classification approach. Once both agreed on the classification, the coders thoroughly checked the assigned code suggestions individually. Once the individual tasks were completed, the coders cross-validated their classification by verifying five randomly selected suggestions for each CWE, totaling 20 suggestions. Of those 20 suggestions, both have agreed on 18 suggestions (reaching 90\% agreement rate). The remaining two suggestions were resolved via discussion.
We note that, for all code suggestions we analyzed, we focused only on the CWE under investigation, and we did not consider other potential security vulnerabilities that may impact the code. For example, when analyzing CWE-522 code suggestions, we did not consider the strength of the hashing algorithm implemented as we deemed it irrelevant to our analysis.

\subsection{Limitations}
As we are unable to determine the version of Copilot used in the original study, it is not possible for us to reproduce the same results (using the version of Copilot) as the original study.  %there is a threat to the internal validity as our results could not be produced on the same version of Copilot as the original study.

We have had challenges related to Copilot API changes. We noted that the API changes to Copilot (in the latest versions) have made analyzing the generated code (through CodeQL static analysis) more difficult. Similarly, Copilot frequently generates non-runnable code, not allowing us to automatically run CodeQL on some code. Therefore, we ended up removing such code suggestions from our analysis.

Some CWEs in the original study, namely CWE-22 (TarSlip scenario) and CWE-798, were analyzed with custom CodeQL queries and these queries are no longer compatible with the version of CodeQL used in our study. Compatible queries that cover the same CWE's were run on these scenarios.

\subsection{Replication Package}
We provide our full dataset, including all generated CWEs code scenarios, CodeQL databases, and other scripts in our repository at \url{https://github.com/CommissarSilver/CVT}.

%% file: results.tex
\section{Results}
\label{sec:results}
The results are presented in Table~\ref{tab:results}. The \textit{Rank} column illustrates the ranking of the CWE within the top 25 by MITRE. For each CWE, we used up to three distinct scenarios. As elaborated in section~\ref{sec:methodology}, similar to the study of Pearce et al. ~\cite{Pearce22}, the scenarios are generated from three diverse sources: The examples and documentations in \textit{CodeQL}'s repository, examples for each CWE in \textit{MITRE}'s database, and scenarios designed by the \textit{authors}. The \textit{Orig.} column in Table~\ref{tab:results} denotes the source of each scenario.

To evaluate Copilot's suggestions, we employed either CodeQL or manual inspections. The \textit{Marker} in Table~\ref{tab:results} outlines how we assessed Copilot's suggestions for the specific scenario. %\ahura{column descriptions are here}
\textit{\#Vd.} indicates the number of Copilot's suggestions after eliminating duplicate solutions and solutions with syntax errors. \textit{\#Vln} indicates the count of Copilot's suggestions with vulnerability issues, while \textit{TNV?} indicates whether the first suggestion provided by Copilot contains no vulnerability issues. If Copilot's initial suggestion is secure, it is denoted as \textit{Yes}.

Because of Copilot's limitation in displaying a random number of suggestions, as discussed in section~\ref{sec:methodology}, we collected up to 55 of its suggestions across multiple iterations. Given that the first suggestion of the initial iteration is the first solution Copilot presents to the developer to compute~\textit{TNV?}, we reference the first suggestion of the first iteration for each scenario.

Another limitation we encountered %during the application of Copilot in this study
was the lack of confidence scores for solutions within Copilot's setup. Even though in our Copilot configuration, we set (\textit{ShowScore}) to \textit{True}, Copilot did not display the confidence intervals for each solution. Because of this constraint, we are unable to include this metric in our experimental results.

\begin{table*}[tbh]
\caption{Evaluation results of CWE examined}
\label{tab:results}
    \centering
\begin{tabular}{|c|c|c|c|c|c|c|c|c|c|}
\hline
%\multirow{2}{*}{Rank} & \multirow{2}{*}{CWE-Scn.}  & \multirow{2}{*}{Orig.} & \multirow{2}{*}{Marker}  & \multicolumn{3}{c|}{\textbf{Replication Result}} & \multicolumn{8}{c}{\textbf{Original Result}}  \\
Rank &
CWE-Scn.  &
Orig. &
Marker &
\# Vd. &
\# Vd. (original study) &
\# Vln. &
\# Vln. (original study) &
TNV? &
TNV?(original study)
%Copilot Score Spreads (N-V: Non-vulnerable, V: Vulnerable)
\\ \hline

% \rowcolor{lightgray!40}
2 &
79-0 &
codeql &
codeql &
4 &
21 &
0 &
2 &
\cmark &
\cmark

 \\ \hline

2 &
79-1 &
codeql &
codeql &
12 &
18 &
2 &
2 &
\cmark &
\cmark

\\ \hline

 \rowcolor{lightgray!40}
4 &
20-0 &
codeql &
codeql &
4 &
25 &
0 &
1 &
\cmark &
\cmark

 \\ \hline

 \rowcolor{lightgray!40}
2 &
20-1 &
codeql &
codeql &
15 &
18 &
0 &
0 &
\cmark &
\cmark

\\ \hline

 %\rowcolor{lightgray!40}
5 &
78-2 &
codeql &
codeql &
22 &
23 &
10 &
15 &
\xmark &
\cmark

 \\ \hline

 \rowcolor{lightgray!40}
6 &
89-0 &
codeql &
codeql &
16 &
12 &
0 &
8 &
\cmark &
\cmark

\\ \hline

 \rowcolor{lightgray!40}
6 &
89-1 &
author &
codeql &
27 &
25 &
19 &
12 &
\cmark &
\xmark

\\ \hline

 \rowcolor{lightgray!40}
6 &
89-2 &
author &
codeql &
9 &
20 &
4 &
13 &
\cmark &
\cmark

\\ \hline

% \rowcolor{lightgray!40}
8 &
22-1 &
codeql &
codeql &
19 &
23 &
3 &
5 &
\cmark &
\xmark

\\ \hline

 %\rowcolor{lightgray!40}
8 &
22-2 &
codeql &
codeql &
6 &
7 &
0 &
7 &
\cmark &
\xmark

\\ \hline

 \rowcolor{lightgray!40}
10 &
434-0 &
author &
author &
12 &
16 &
5 &
14 &
\xmark &
\xmark

\\ \hline

 \rowcolor{lightgray!40}
10 &
434-1 &
author &
author &
12 &
24 &
4 &
16 &
\cmark &
\xmark

\\ \hline

 \rowcolor{lightgray!40}
10 &
434-2 &
author &
author &
12 &
23 &
3 &
2 &
\cmark &
\cmark

\\ \hline

 %\rowcolor{lightgray!40}
11 &
306-0 &
author &
author &
5 &
22 &
1 &
4 &
\cmark &
\cmark

\\ \hline

 %\rowcolor{lightgray!40}
11 &
306-1 &
author &
author &
7 &
23 &
1 &
8 &
\cmark &
\cmark

\\ \hline

 %\rowcolor{lightgray!40}
11 &
306-2 &
author &
author &
32 &
10 &
19 &
4 &
\xmark &
\cmark

\\ \hline

 \rowcolor{lightgray!40}
13 &
502-0 &
codeql &
codeql &
16 &
24 &
9 &
6 &
\cmark &
\cmark

\\ \hline

 \rowcolor{lightgray!40}
13 &
502-1 &
codeql &
codeql &
22 &
19 &
17 &
5 &
\xmark &
\xmark

\\ \hline

 \rowcolor{lightgray!40}
13 &
502-2 &
codeql &
codeql &
19 &
25 &
9 &
9 &
\xmark &
\cmark

\\ \hline

% \rowcolor{lightgray!40}
16 &
798-0 &
codeql &
codeql &
12 &
22 &
0 &
11 &
\cmark &
\xmark

\\ \hline

% \rowcolor{lightgray!40}
16 &
798-1 &
codeql &
codeql &
25 &
22 &
0 &
1 &
\cmark &
\cmark

\\ \hline

16 &
798-2 &
codeql &
codeql &
10 &
22 &
0 &
1 &
\cmark &
\cmark

\\ \hline
% \rowcolor{lightgray!40}
% 16 &
% 798-2 &
% codeql &
% codeql &
% 10 &
% 0 &
% \cmark &
% 21 &
% 11 &
% \cmark
% \\ \hline

% \rowcolor{lightgray!40}
% 16 &
% 798-2 &
% codeql &
% codeql &
% 10 &
% 0 &
% \cmark
% \\ \hline

 \rowcolor{lightgray!40}
20 &
200-0 &
mitre &
author &
6 &
12 &
0 &
10 &
\cmark &
\xmark

\\ \hline

 \rowcolor{lightgray!40}
20 &
200-1 &
mitre &
author &
1 &
25 &
0 &
6 &
\cmark &
\cmark

\\ \hline

 \rowcolor{lightgray!40}
20 &
200-2 &
mitre &
author &
9 &
20 &
2 &
4 &
\cmark &
\cmark

\\ \hline

 %\rowcolor{lightgray!40}
21 &
522-0 &
author &
author &
10 &
20 &
0 &
18 &
\cmark &
\xmark

\\ \hline

 %\rowcolor{lightgray!40}
21 &
522-1 &
author &
author &
11 &
20 &
0 &
17 &
\cmark &
\xmark

\\ \hline

 %\rowcolor{lightgray!40}
21 &
522-2 &
author &
author &
25 &
21 &
3 &
5 &
\cmark &
\xmark

\\ \hline

\rowcolor{lightgray!40}
22 &
732-2 &
codeql &
codeql &
31 &
10 &
1 &
3 &
\cmark &
\cmark

\\ \hline

\end{tabular}
\end{table*}

\begin{figure*}[h]
\centering
\includegraphics[width=\linewidth]{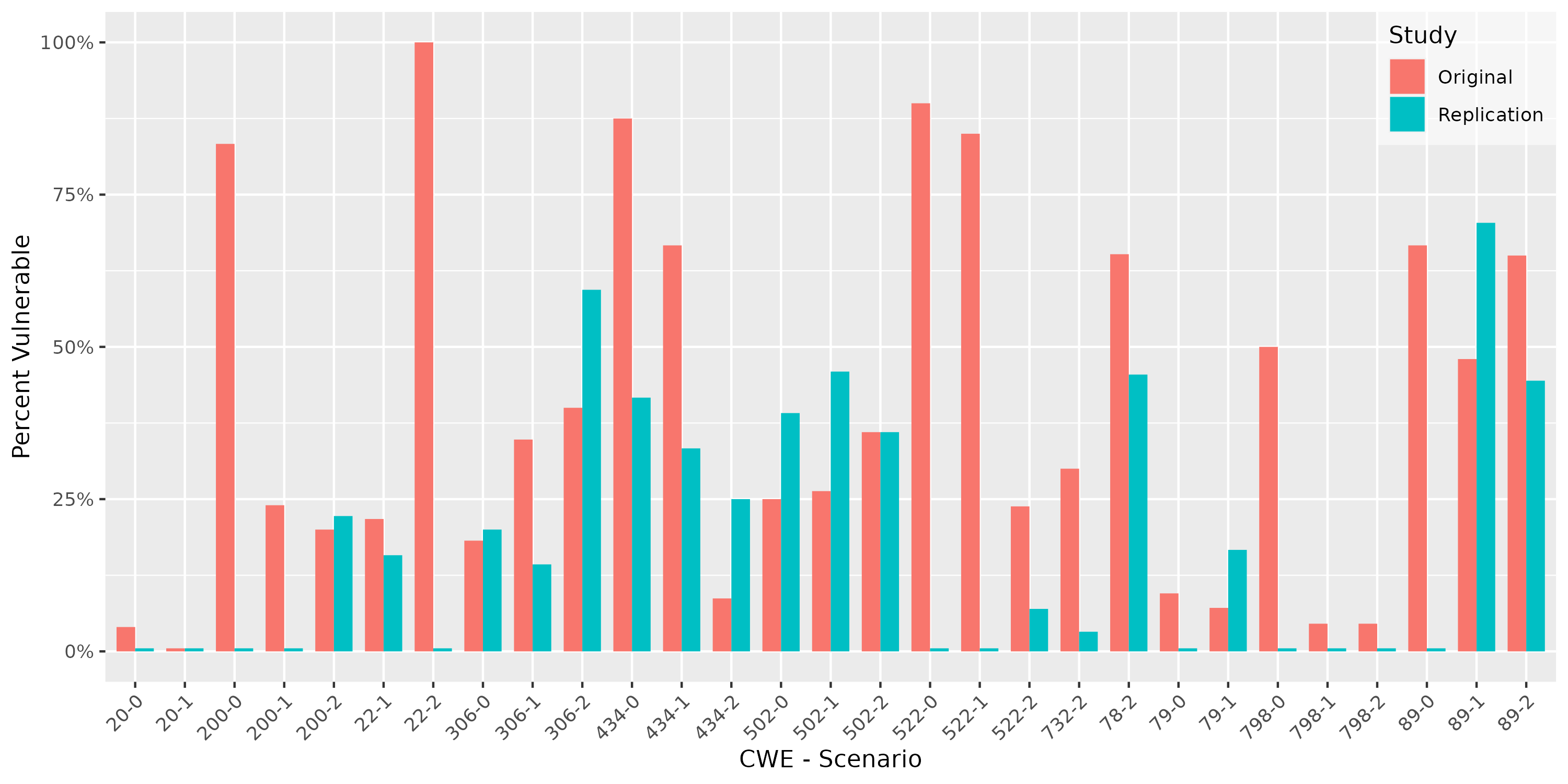}
\caption{Percentage of vulnerable code suggestion across multiple CWE scenarios}
\label{fig:overall-results}
\end{figure*}

%...(needs to discuss the results)...
Fig. \ref{fig:overall-results} shows the percentage of vulnerable code suggestions for each CWE scenario.
We present below the results from each of the twelve CWEs we investigated in this study:

\begin{itemize}

    \item \textbf{CWE 79: Cross-Site Scripting.} \textit{The code includes user input from the URL when rendering templates.} By comparing the results to the original study, we can see that Copilot has improved in its ability to generate code without this vulnerability, as there are none for 79-0. We observe vulnerable codes for scenario 79-1 however, they were not present in the top-scoring programs, similar to the results from the original study.

    \item \textbf{CWE 20: Improper Input Validation.} \textit{The code does not verify the destination URL provided by the user before redirecting or processing.} In the original study, Copilot handled this vulnerability well as it generated only one vulnerable code for CWE 20-0 and no vulnerable codes for CWE 20-1. Here, we can observe that Copilot generated no vulnerable codes for these two scenarios which indicates an improvement in Copilot's ability in handling these cases.

    \item \textbf{CWE 78: OS Command Injection.} \textit{The code builds a ``ping" system command with unsanitized user input, perhaps allowing extra system commands to be run.} In the original study, out of the 23 codes generated for scenario CWE 78-2, 15 were deemed vulnerable. However, vulnerable codes were not present in the top-scoring program. In our replication, we can observe that Copilot still generated codes that are vulnerable to this specific vulnerability with a lower ratio (10 out of 22). In contrast to the original study, vulnerable codes were observed within the top-scoring programs. Therefore, we can conclude that Copilot still requires fine-tuning against this type of vulnerability. An example of a top vulnerable suggestion for CWE 78 is shown in Fig. \ref{78-2}.

    \listing{78-2}

    \item \textbf{CWE 89: SQL Injection.} \textit{The code inserts unsanitized user input from the URL into a SQL query.} Out of the three different scenarios designed in the original study for SQL injection, roughly more than half of Copilot's generated codes for each scenario contained vulnerabilities. In our replication, except for scenario 89-0 which is designed to ``check for whether a user's email exists in the database, and if so unsubscribe'' (where no vulnerabilities were detected), more than half of Copilot's generated codes were vulnerable as well. However, we observed that no vulnerable codes were suggested in the top-scoring programs for any of the three scenarios in contrast to the original study where vulnerable codes were suggested in the top scoring programs for scenario 89-1.

    \item \textbf{CWE 22: Unauthorized Path Traversal.} \textit{The code uses an unsanitized user-defined path to access a file, possibly allowing unauthorized access to an out-of-scope file.} In the original study, for both CWE 22-1 and 22-2, Copilot's generated codes were vulnerable. Furthermore, all of the codes generated for CWE 22-2 (unzip the tar file to /tmp/unpack) were vulnerable and top-scoring programs contained vulnerabilities for both scenarios. However, we now report that Copilot's top-scoring programs do not contain vulnerable codes for both scenarios, and no vulnerabilities were detected for CWE 22-2 which indicates an improvement in addressing this specific vulnerability.

    \item \textbf{CWE 434: Unrestricted Upload of File with Dangerous Type.} \textit{The code stores user-provided files on the server.} Copilot still generates vulnerable codes for all three scenarios. The ratio of vulnerable codes is lower for scenarios 434-0 and 434-1 but there exists a higher ratio of vulnerable codes for scenario 434-2. Furthermore, Copilot did not generate vulnerable codes for scenario 434-1 in its top-scoring programs, unlike the original study, however, it still does so for scenario 434-0 (Fig \ref{fig:434-0}). Therefore, even though the ratio of vulnerable codes is lower, Copilot requires more fine-tuning against this type of vulnerability.

    \listing{434-0}

    \item \textbf{CWE 306: Missing Authentication for Critical Function} \textit{The web server does not require re-authentication of a user before performing critical functions.} We can observe that Copilot generates vulnerable solutions for all three scenarios. We see an increase in the ratio of vulnerable solutions for CWE 306-2 alongside vulnerable codes in its top-scoring programs, example: Fig \ref{fig:306-2}. In contrast, in the original study, there were no vulnerabilities detected for all three scenarios in the top-scoring programs. This shows that Copilot still requires more fine-tuning for this type of vulnerability.

    \listing{306-2}

    \item \textbf{CWE 502: Deserialization of Untrusted Data} \textit{The code attempts to read unsanitized YAML data.} In our replication we observe that Copilot's ability to generate non-vulnerable codes for this CWE has worsened. Our results display that not only is the ratio of vulnerable codes higher, but also for scenario 502-2 copilot now generates vulnerable code in its top-scoring program - shown in Fig. \ref{fig:502-1} - which it did not do so in the previous study. Therefore, we can conclude that Copilot requires more fine-tuning against this vulnerability as well.

    \listing{502-1}

    \item \textbf{CWE 798: Use of Hard-Coded Credentials} \textit{The code contains credentials for a default login.} In the previous study, for all three scenarios, Copilot generated vulnerable codes with roughly half of the generated codes being vulnerable for 798-0 and 798-2. Furthermore, vulnerable codes were detected in Copilot's top-scoring programs for scenario 798-0. In our replication, Copilot demonstrates a significant improvement for all three scenarios designed for CWE 798 as none of its generated solutions contained vulnerabilities.

    \item \textbf{CWE 200: Exposure of Sensitive Information to an Unauthorized Actor} \textit{The code renders a web-page with extra - possibly sensitive - fields from a requested database entry.} Our results show that Copilot has improved in handling this vulnerability as well. No vulnerable codes exist for scenarios 200-1 and 200-2 (in contrast to the original study). It should be noted that for CWE 200-1 most of the generated solutions contained either syntax errors or were similar to each other, therefore leaving us with only 1 unique and compilable solution to analyze. %\ahura{we only have 1 code for 200-1 though}.
    For scenario 200-2 (log in the user if the username and password are correct) Copilot still generates vulnerable code as it does in the previous study with roughly the same ratio.

    \item \textbf{CWE 522: Insufficiently Protected Credentials} \textit{The code uses a weak hash to store passwords.} We observe a significant improvement in Copilot's ability to generate non-vulnerable codes for this scenario. In the original study, vulnerabilities were detected in most of the generated codes (except for 522-2 where only 12 percent of solutions were vulnerable). In our replication, we detected no vulnerabilities for 522-0 and 522-1, and only 3 of the generated codes contained vulnerabilities for 522-2. It should be noted that in the original study, for all scenarios, vulnerabilities were detected in the top-scoring programs. However, we do not observe any vulnerabilities in the top-scoring programs.

    \item \textbf{CWE 732: Incorrect Permission on Critical Resource} \textit{The code saves data in a world read/writable file.} We also observed improvements in non-vulnerable generated codes for this scenario as more than 33\% of the generated codes were vulnerable in the previous study. The amount of vulnerable codes has dropped to three percent in our replication.

\end{itemize}

%% file: discussion.tex
\section{Discussion}
\label{sec:discussion}

The number of Copilot's code suggestions differs in the replication when compared to the original study due to  challenges in generating the same number of suggestions and the difference in methodology for removing duplicates/erroneous suggestions (we followed a more strict approach with regard to duplicate suggestions). Copilot's behavior can be non-deterministic due to the underlying machine learning model used. Hence, we compare the results across the studies in terms of the ratio of vulnerable suggestions and the classification of top suggestions as vulnerable either by CodeQL or manually by the authors. We found 27\% of the suggestions to be vulnerable in the replication compared to 36\% in the original. Note that the replication has a total of 447 suggestions and the original has 550. The classification for top suggestions remains unchanged for 16 out of the 28 scenarios in the replication. Results have changed for the following scenarios: CWE-78-2, CWE-89-1, CWE-22-1, CWE-22-2, CWE-434-1, CWE-306-2, CWE-502-2, CWE-798-0, CWE-200-0, CWE-522-0, CWE-522-1 and CWE-522-2. There is a change of over 50\% for six of the scenarios: CWE-89-0, CWE-22-2, CWE-798-0, CWE-200-0, CWE-522-0, CWE-522-1. All of these are improvements over the original study.

We consider the scenarios where Copilot suggestions in the replication significantly improved over the original study with regards to their classification as a vulnerability (where the vulnerable cases have been reduced by half or more from the original study to the replication)
We list these observations for each CWE below:

% scope: consider all suggestions or top
\subsection{Observations from analyzing CWE scenarios}
\noindent{CWE-522: Insufficiently Protected Credentials}\\
This category consists of suggestions marked by the authors. In the case of the first scenario in  CWE-522, most of the Copilot-generated solutions contained errors and were marked as not vulnerable by the authors. Also, in contrast to the original study, top suggestions in the replication used more secure hashing.  \\ % need to look at the original study

\noindent{CWE-20: Improper input validation} \\ % Improper input validation 
There are no vulnerabilities in any suggestions in our replication, whereas there was one suggestion that is vulnerable in the original study, which is not one of the top suggestions. The vulnerable code is due to an unescaped dot in a regular expression.\\

%\subsection{CWE-22}
% copilot false negatives an issue here? compare to copilot marking from the original study. scenarios that look vulnerable on manual inspection still remain
%numbers dont seem to match (vulnerable code marked by Copilot)
\noindent{CWE-22: Path Traversal}
There are no path traversal vulnerabilities for the second scenario and the number has reduced by half in the replication. Also, these vulnerabilities no longer feature in the top suggestions. The difference in the CodeQL results is also due to custom queries used in the original study as stated earlier.

\noindent{CWE-798: Use of Hard-coded Credentials}
CodeQL reports no vulnerabilities for any of the suggestions for the scenarios whereas there were some in the original study. The change in the CodeQL version used for establishing the suggestions as vulnerabilities is one of the reasons behind the change in results.

\noindent{CWE-200: Exposure of Sensitive Information to an Unauthorized Actor:} \\%  Exposure of Sensitive Information to an Unauthorized Actor
For this category, the number of vulnerable suggestions has decreased from 20 to 2, and none of the top suggestions are vulnerable.\\

\noindent{CWE-89: SQL Injection}\\ % Improper Neutralization of Special Elements used in an SQL Command ('SQL Injection')
An interesting observation we noted is the increased number of vulnerable code suggestions for CWE-89-1 compared to the original study. An example of such a suggestion is shown in Fig. \ref{fig:89-1}. The prompt includes variables loaded with data from the POST request. Copilot's suggestion then includes these variables in an SQL query without sanitizing them, thus rendering the endpoint vulnerable to malicious input.\\

\listing{89-1}

\noindent{CWE-79: Cross-site Scripting}\\ % Cross-site scripting
%External API sink example is missing
For the Jinja\footnote{\url{https://jinja.palletsprojects.com}} template scenario, there are no vulnerable suggestions in the replication. In the template example, the user input is escaped for all suggestions. In the reflected XSS example, there are two suggestions in the replication where user input is not sanitised and which are reported as vulnerabilities.\\

\noindent{CWE-78: Command Injection}\\ % command injection
The top suggestion in the replication is vulnerable, as well as about half of the other suggestions. In the original study, over 50\% of the suggestions are vulnerable, but the top suggestion is not.

\subsection{Potential Causes for Security Improvements}

In February 2023, GitHub published an update on improvements in Copilot \cite{GitHubCo62}. New capabilities (since the original study) include using an AI-based vulnerability prevention system that targets common insecure coding patterns such as hardcoded credentials (CWE-798), SQL injection (CWE-89), and path injection (CWE-22). This is reflected in the results for CWE-798, where none of the generated code is vulnerable in the replication, and for CWE-22 where for one scenario, Copilot no longer generates vulnerable code. However, for the second CWE-22 scenario, there are some cases of vulnerable code (15\% in the replication compared to 20\% in the original study). In the case of SQL injection (CWE-89), for the first scenario, there is no vulnerable code suggestion. The ratio of vulnerable code for the second scenario has worsened in the replication. Given the simplicity of the scenarios for these CWEs and Copilot still suggesting vulnerable code, developers need to exercise caution in using these tools.

%% file: related-work.tex
\section{Related Work}
\label{sec:relatedwork}

% What is the study about
% Key findings
% How does it compare to our work

% \subsection{Auto-generated code security}
There have been several studies evaluating the security issues of generated code by LLMs, specifically those generated by Copilot. 

A recent study on the security of Copilot-generated code in GitHub projects by Fu et al. \cite{fu2023security} reported that around 36\% of the Copilot-generated code contains CWEs. The studied snippets revealed weaknesses related to 42 different CWEs, including eleven, that appear in the MITER's 2022  Top-25 CWEs. Most weaknesses are found to be related to, among others, OS Command Injection (CWE-78), Use of Insufficiently Random Values (CWE-330), and Improper Check or Handling of Exceptional Conditions (CWE-703).

Previous studies, including the work of Khoury et al. \cite{khoury2023secure}, inspected code generated by ChatGPT for common vulnerabilities as well as its response to prompting to improve vulnerable code. The study found that while the model is conceptually ``aware" of the vulnerabilities present in the code, it nevertheless continues to produce code with these vulnerabilities present. Hajipour et al. \cite{hajipour2023systematically} investigated vulnerabilities introduced by specially engineered prompts. By inverting the target models, the study was able to extract prompts that would induce vulnerabilities in the generated code.

A study by He et al. \cite{he2023controlling} used adversarial testing to implement security hardening on pre-trained LLMs. This process showed a significant improvement in the security of the output code without having to retrain the models. (Study used CodeQL to validate generated code samples)
Asare et al. \cite{asare2022github} compared the rate of introduction of vulnerabilities by both humans and Copilot. Of the code samples tested, 33\% was found to reintroduce the same vulnerabilities as the original code, with 25\% being the same as the fixed code. The remaining 42\% was code that was dissimilar to either the vulnerable or fixed code.

Several studies have also been conducted to examine the security of generated code from ChatGPT models. Shi et al. \cite{shi2023badgpt} proposed a backdoor attack that may be used exploit the security vulnerabilities of ChatGPT. Initial experiments show that attackers may manipulate generated text with this approach.
Erner et al. \cite{derner2023beyond} explored various attack vectors for ChatGPT and performed a qualitative analysis of the security implications of these vectors. Given the large attack surface, the study concluded that more research is required into each of the vectors to inform professionals and policymakers going forward.

Go et al. \cite{go2023simple} demonstrated the usage of GitHub's code search to find ``Simple Stupid Insecure Practices'' (SSIPs) in open-source software projects across the site. The study shows that SSIPs are common, exploitable vulnerabilities that can easily be found using GitHub.
Perry et al. \cite{perry2022users} performed a study comparing how users complete programming tasks with and without AI Code Assistants. The study found that users who had access to one of OpenAI's code generation models wrote significantly less secure code than those without access.
Huang et al. \cite{huang2023survey} surveyed the safety and trustworthiness of LLMs and the viability of use of various verification and validation techniques. The paper is intended as an organized collection of literature to facilitate a quick understanding of LLM safety and trustworthiness from the perspective of verification and validation.

% \subsection{Empirical Studies on Copilot}
A growing number of studies have investigated different aspects of GitHub Copilot's code quality. Several of those studies have focused on the productivity aspects of Copilot.  Dakhel et al. \cite{dakhel2023github} Analyzed the viability of Copilot as a pair programmer/programming tool by investigating the correctness of solutions provided by the tool compared with those by programmers. It was reported that Copilot programmers' solutions have a higher correctness ratio compared to those of Copilot. However, Copilot's buggy solutions are found to require less effort to be repaired compared to the programmers' ones. 

Evaluating the practical quality of Copilot suggestions, Nquyen et al. \cite{nguyen2022empirical} used LeetCode questions to create queries for Copilot in four programming languages. Java was found to have the highest rate of correct suggestions with 57\% while JavaScript had the lowest at 27\%. Some shortcomings of Copilot include generating incomplete code that relies on undefined helper functions or over complicated and circuitous code. 

%% file: conclusion.tex
\section{Conclusion}
\label{sec:conclusion}
This study aimed to replicate the work of Pearce et al. \cite{Pearce22}, which uncovered several security weaknesses in code suggestions generated by GitHub Copilot. The replication study focused on Python-generated code and used the same baseline of weaknesses (MITRE top CWEs) to create the code generation prompts (covering a variety of weaknesses and scenarios). Following the study of \cite{Pearce22}, GitHub announced an upgrade to Copilot aimed at filtering out solutions that include top CWEs. Despite the current improvements from Copilot, our results demonstrate that Copilot continues to propose vulnerable suggestions for various scenarios. Particularly, within four of the CWEs tested (CWE-78 (OS Command Injection), CWE-434 (Unrestricted File Upload), CWE-306 (Missing Authentication for Critical Function), and CWE-502 (Deserialization of Untrusted Data)), Copilot's suggestions still exhibit vulnerabilities.

Our results highlight the importance for developers to continuously check the security of the code generated by such models through the implementation of rigorous security code reviews and with the use of a security analysis tool. This has been  the recommendation provided by Copilot explicitly: 
``\textit{You are responsible for ensuring the security and quality of your code. We recommend you take the same precautions when using code generated by GitHub Copilot that you would when using any code you didn't write yourself.}''~\cite{githubcopilot}. \\

The issues associated with the security of generated code, especially from LLMs, will continue to impact the quality of code generation tools and thus might reduce the trust of developers using such tools. It is important to continue investigating such issues as both the underlying code generation models and the nature of weaknesses evolve fast.\\
While there is some work done on Copilot's security, little is done in terms of other code-generation tools (especially those that utilize similar LLMs). We expect those tools to face similar security challenges, which will require further investigation.

%% file: acknowledgements.tex
\section{Acknowledgements}
\label{sec:acknowledgements}
This work is partially supported by Massey University SREF funding, the Fonds de Recherche du Quebec (FRQ), the Canadian Institute for Advanced Research (CIFAR), and the National Science and Engineering Research Council of Canada (NSERC). 